\documentclass[11pt]{article}
\input{glyphtounicode}\pdfgentounicode=1
\usepackage[T1]{fontenc}
\usepackage[utf8]{inputenc}
\usepackage[letterpaper,top=.91in,bottom=.94in,left=1.10in,right=1.10in]{geometry}
\usepackage{amsmath,amssymb,amsthm,lmodern,newtxtext,microtype,booktabs}
\usepackage{graphicx,tikz,pgfplots,xcolor,enumitem}
\pgfplotsset{compat=1.18}
\usepackage[round,authoryear]{natbib}
\usepackage{xurl}
\usepackage{hyperref}
\setcitestyle{authoryear,round,aysep={},citesep={;}}
\colorlet{linkblue}{blue!70!black}
\hypersetup{colorlinks=true,linkcolor=linkblue,citecolor=linkblue,urlcolor=linkblue,pdftitle={When Safety Warnings Accelerate Rivals},pdfauthor={Haotian Ding},pdfsubject={Voluntary disclosure and competition among AI developers}}
\usepackage{caption}
\newtheorem{theorem}{Theorem}
\newtheorem{proposition}{Proposition}
\newtheorem{corollary}{Corollary}
\newcommand{\Prob}{\mathop{\rm Pr}\nolimits}
\newcommand{\ind}{\mathbf{1}}
\newcommand{\Pos}[1]{[#1]_{+}}
\setlist[itemize]{nosep,topsep=4pt}
\title{\vspace{-.45in}\bfseries When Safety Warnings Accelerate Rivals}
\author{\normalsize Haotian Ding\\[2pt]
\small School of Finance, Renmin University of China, Beijing, China\\
\small \href{mailto:dht6188@ruc.edu.cn}{dht6188@ruc.edu.cn}}
\date{\small November 2026}
\begin{document}
\maketitle
\vspace{-1.1em}
\begin{abstract}\microtypesetup{protrusion=false}\noindent
A frontier artificial intelligence (AI) laboratory can warn defenders about a dangerous capability only after observing verifiable evidence of it, yet the same warning can indirectly signal research progress to a competitor. We study voluntary disclosure when a verified warning triggers defensive mitigation while a rival chooses whether to bring forward its next research investment. In the unique active equilibrium, whenever hazard evidence predicts a sufficiently strong competitive threat, a warning induces pursuit with positive probability despite certifying the hazard. Because pursuit penalizes advanced laboratories more heavily, they withhold reports more often. In the mixed equilibrium, raising a reporting subsidy increases both the volume of warnings and the probability that each warning triggers acceleration, yet leaves the rival's posterior belief about the sender's progress completely unchanged; observing stable technical composition among published reports therefore does not imply unchanged competitive behavior. Adding a narrower confidential channel alongside public warnings can likewise slash per-warning pursuit while leaving the public-warning posterior unchanged. Finally, if mitigating a shared accident is the sender's sole return to reporting, revealed preference implies that voluntary warnings strictly lower expected accident risk despite inducing pursuit. External rewards can overturn this guarantee: across admissible subsidies, minimizing accident risk calls for a strictly positive---yet capped---subsidy only when the sender's private accident stake is sufficiently small.
\end{abstract}

\section{Introduction}
On April 7, 2026, Anthropic disclosed that an unreleased model, Claude Mythos Preview, had identified serious vulnerabilities in widely used software, restricting general access while launching Project Glasswing to help defenders patch affected systems \citep{anthropic}. The announcement informed the public both that the hazard warranted protective action and that its developer had built a model capable of discovering it. OpenAI's chief executive subsequently characterized the presentation as \emph{fear-based marketing} in an interview \citep{core}---a remark that captures a natural interpretation available to a competitor: a safety warning can simultaneously advertise a technical advance.

Why would a laboratory publish verifiable evidence of danger if doing so could accelerate a rival's research? The tension arises because a public warning reaches two audiences: defenders who reduce the likelihood of an accident, and a competitor eager to learn whether it is falling behind. Even a laboratory motivated solely by avoiding a common accident may accept the competitive cost of pursuit in order to secure defensive mitigation. Crucially, the choice to speak shapes what the rival infers. Because laboratories holding the most valuable contested leads suffer the steepest losses when a competitor speeds up, they have the strongest incentive to withhold positive findings. A rational rival accounts for this selection and discounts the progress implied by a voluntary warning, even while accepting the verified hazard claim without reservation.

Recent industry episodes illustrate the institutional setting while highlighting an important distinction between direct and indirect capability revelation. Anthropic's April announcement paired its warning with capability benchmark summaries while routing unpatched vulnerability details through coordinated disclosure, so readers learned about model performance partly directly; shortly afterward, OpenAI expanded its own cybersecurity trusted-access program, which had operated since February \citep{openfeb,openapr}. While public records cannot establish whether Anthropic's disclosure altered OpenAI's release schedule---nor isolate indirect inference from direct benchmark reporting---a separate episode documents competitive responsiveness to progress signals: \citet{opensep} reports launching a targeted mathematical effort on the Navier--Stokes problem within days of hearing rumors of an external breakthrough. Our model strips away direct capability reporting so that a warning certifies \emph{only} the hazard, isolating when correlation and selective withholding alone supply the progress news that triggers pursuit.

Specifically, a laboratory privately observes a binary hazard, its own research progress, and a reporting cost before deciding whether to disclose verifiable evidence of the hazard. Observing the disclosure decision, a competitor chooses between its planned pace and an accelerated investment. Confirming the hazard raises the competitor's private risk cost of accelerating, whereas inferring that the sender has advanced raises the commercial value of catching up. In the unique active equilibrium, a voluntary warning induces pursuit with positive probability if and only if a randomly sampled positive evaluation would prompt acceleration. Pursuit remains probabilistic across a wide parameter range even though every technical claim is verified: if the rival always accelerated after a warning, too many advanced laboratories would stay silent for pursuit to remain worthwhile.\footnote{When the private return to reporting is small, a silent equilibrium coexists with active disclosure in unrefined perfect Bayesian equilibrium and survives the Intuitive Criterion of \citet{chokreps}, but fails Criterion D1 because low-progress senders gain from breaking silence across a strictly wider set of rival responses.}

Dual-audience disclosure and indirect signaling have precedents in accounting and industrial organization \citep{wagenhofer,heinle}. Embedding \emph{defensive mitigation} and \emph{exposure to a shared accident} into this structure generates three distinctive predictions. First, in the mixed equilibrium, raising a reporting subsidy increases both the number of warnings and the probability that each warning triggers acceleration, while leaving the rival's posterior belief about the sender's progress fixed. Holding rival behavior constant, higher rewards draw reluctant advanced laboratories into the reporting pool; in equilibrium, the rival accelerates more frequently by just enough to offset that entry and restore indifference. Consequently, an observer who tracks only report volume and the technical composition of published warnings can miss the induced competitive escalation entirely. Second, introducing a narrower confidential reporting channel alongside public disclosure induces advanced reporters to split across channels, sharply reducing pursuit per public warning while again leaving the progress posterior of public warnings unchanged. Third, when mitigating the common accident is the sender's sole return to reporting, its willingness to bear verification costs and rival pursuit reveals that mitigation outweighs pursuit-induced accident risk. External subsidies break that revealed-preference discipline, creating a non-monotone safety trade-off between drawing out withheld warnings and provoking pursuit.

\section{Related literature}
Our analysis builds on two strands of disclosure theory. The first studies voluntary disclosure when a single message reaches both a target audience and a strategic competitor. \citet{wagenhofer} analyzes a firm that discloses verifiable private information to capital markets while a rival chooses whether to take an adverse binary action, showing how proprietary costs can sustain partial-pooling equilibria. Closer to our informational structure, \citet{heinle} show theoretically and empirically that verifiable short-term earnings forecasts indirectly signal correlated long-term prospects, attracting rival investment; in their setting, firms with stronger long-term prospects have greater incentives to issue forecasts despite competitive costs. In R\&D competition, \citet{bhatt} and \citet{jansen} likewise examine disclosures that leak technically or commercially relevant information to rivals. Here, a warning contains neither technical instructions nor direct performance metrics, and defensive mitigation rewards certifying the hazard itself rather than signaling high project value. Because rival pursuit penalizes advanced senders more heavily ($K_H>K_L$), selection runs \emph{against} advanced types: high-progress laboratories withhold verifiable findings more often \citep[cf.][]{verrecchia}. Combining this downward selection with a rival mixing over pursuit yields our central comparative static: subsidies and parallel confidential channels shift rival pursuit rates while pinning the progress content of public warnings at the rival's indifference threshold.

A second literature examines information revelation in R\&D and patent races. \citet{baik} study duopolists that decide ex ante whether to commit to disclosing R\&D investments, finding mutual commitment to transparency in quantity and price games. Focusing instead on ex post voluntary announcements of interim breakthroughs, \citet{kocourek} characterize when revealing progress discourages a rival from continuing research, and \citet{shao} study milestones that signal both feasibility and a wider lead, showing that firms disclose when deterrence prevails. In our model, defensive mitigation supplies an independent benefit from speaking, making laboratories willing to warn even when disclosure strictly \emph{accelerates} rather than deters the rival.

The direction of rival investment and the shared exposure to industry-wide harm also connect our work to risk regulation and AI race dynamics. \citet{august} show that sharing cyber-threat signals can induce rivals with low priors to invest more in operational resilience. In our framework, external defenders undertake protection while the rival decides whether to rush capability research that heightens accident risk; a single warning can thus reduce risk via mitigation and raise it via pursuit. Crucially, because both firms suffer losses in an industry accident, the sender's participation constraint yields a revealed-preference safety guarantee whenever external reporting rewards are absent. This complements \citet{friehe}, who study how mandatory versus voluntary product-risk disclosure shapes endogenous testing, and \citet{obrien}, who propose coordinated confidential disclosure institutions for frontier AI. Finally, whereas dynamic models of technology and AI races highlight how exogenous observability gaps shape risk-taking and restraint \citep{emery,biais,fudenberg}, we endogenize observability by letting the laboratory holding hazard evidence choose whether and through which channel to speak.

\section{Environment}
Consider an interim safety evaluation and a competitor's subsequent decision over research timing. Laboratory $A$ privately observes its research progress $t\in\{L,H\}$ and a binary hazard state $r\in\{0,1\}$. State $r=1$ indicates that the evaluation has generated verifiable evidence of a dangerous capability, whereas state $H$ indicates that $A$ has reached a technical milestone that still leaves its competitor, Laboratory $B$, a commercially contestable opportunity to catch up. Let
\[
p=\Prob(r=1),\qquad q=\Prob(H\mid r=1),\qquad q_0=\Prob(H\mid r=0),
\qquad 0<p<1,\quad 0<q_0<q<1.
\]
The inequality $q>q_0$ captures a natural property of capability evaluations: dangerous findings emerge more frequently in more advanced projects, though they can occur at either stage of progress.

Laboratory $A$ also draws a private reporting cost $k\sim U[0,C]$, independent of $(r,t)$, reflecting expenses such as independent validation, redaction of sensitive exploit details, or coordination overhead. When $r=1$, $A$ chooses between issuing a public warning $W$ and remaining silent $S$; when $r=0$, verifiable evidence is absent and only $S$ is feasible. A warning certifies $r=1$ without directly revealing $t$. Upon observing $W$, external defenders implement a prescribed protective measure, such as patching vulnerable software or tightening access filters. Because patching addresses the concrete vulnerability documented in $W$, its efficacy depends on $r=1$ rather than on $A$'s broader commercial progress $t$.

Observing whether a warning was issued, Laboratory $B$ chooses between maintaining its planned research schedule ($a=0$) and accelerating its next effort ($a=1$), such as bringing forward a training run or reallocating compute and engineering talent to an immediate capability push. Focusing on $B$'s timing choice after $A$'s evaluation isolates the informational spillover of safety warnings without confounding it with simultaneous signaling by both firms.

\paragraph{The competitor.} Laboratory $B$'s incremental payoff from choosing $a=1$ rather than $a=0$ in state $(r,t)$ is
\begin{equation}
 D_{rt}=v_t-c-hr,\qquad v_H>v_L,\quad h>0. \label{eq:D}
\end{equation}
Here $v_t$ represents the commercial gain from earlier investment---for instance, the incremental probability of capturing a contested market or contract multiplied by its value. When $A$ has reached $H$, moving early is more valuable to $B$ while the market remains contestable ($v_H>v_L$).\footnote{If $A$'s lead in state $H$ were instead insurmountable, $v_H<v_L$ would hold and progress news would discourage $B$'s investment, reversing selection toward the deterrence motive in \citet{kocourek} and \citet{shao}.} The constant $c$ captures direct acceleration costs plus baseline risk costs, while a confirmed hazard ($r=1$) adds $h>0$ to the expected risk cost of rushing ahead.

Let $D_0=v_L+(v_H-v_L)q_0-c$ and $D_1=v_L+(v_H-v_L)q-c-h$ denote $B$'s expected incremental payoffs conditional on $r=0$ and $r=1$, respectively, prior to any disclosure selection. We maintain
\begin{equation}
D_0<0,\qquad \bar D=(1-p)D_0+pD_1<0,\qquad
0<u\equiv\frac{c+h-v_L}{v_H-v_L}<1. \label{eq:domain}
\end{equation}
Under \eqref{eq:domain}, $B$ prefers its planned pace in the absence of news ($\bar D<0$) and conditional on no hazard ($D_0<0$). Once $r=1$ is certified, accelerating yields a nonnegative expected payoff if and only if $B$'s posterior belief on $t=H$ is at least $u$.

\paragraph{The reporting laboratory.} Conditional on $r=1$, $A$'s payoff (up to state-dependent constants) is
\begin{equation}
 U_A=(b-k)\ind_{\{W\}}-K_t a,\qquad
 b>0,\quad 0<K_L<K_H,\quad C>b+K_H . \label{eq:A}
\end{equation}
The parameter $b$ captures $A$'s gross private return to reporting, encompassing the benefit of defensive mitigation as well as any external reporting reward or subsidy. Rival acceleration imposes a loss $K_t$ on $A$ that is strictly larger when $A$ has an advanced lead at stake ($K_H>K_L$). The upper support condition $C>b+K_H$ ensures interior reporting probabilities ($z_t<1$) even if a warning were to deter acceleration, while allowing $b\le K_L$ accommodates environments in which fear of pursuit can silence the entire industry.

Our baseline solution concept is perfect Bayesian equilibrium (PBE). Both laboratories maximize expected payoffs at every information set, beliefs satisfy Bayes' rule on the equilibrium path, and verifiability restricts off-path beliefs following $W$ to assign probability one to $r=1$. An equilibrium is \emph{active} if a strictly positive measure of evaluations results in a public warning.

\section{Equilibrium}
Let $a_W,a_S\in[0,1]$ denote $B$'s probabilities of accelerating after $W$ and after $S$, respectively, and define the net pursuit differential $y=a_W-a_S$. Conditional on $r=1$, Laboratory $A$ prefers $W$ to $S$ when $k\le b-K_t y$, yielding type-specific reporting probabilities $z_t(y)=\Pos{b-K_t y}/C$, where $[w]_+=\max\{0,w\}$. Whenever warnings occur with positive probability, Bayes' rule implies that the posterior probability of advanced progress following $W$ is
\begin{equation}
 m(y)=\frac{q z_H(y)}{qz_H(y)+(1-q)z_L(y)}. \label{eq:posterior}
\end{equation}
Because endogenous reporting decisions enter both the numerator and the denominator of \eqref{eq:posterior}, observed warnings systematically differ in composition from the underlying population of positive evaluations.

\begin{theorem}[Disclosure and pursuit]\label{th:main}
Under \eqref{eq:domain}--\eqref{eq:A}, there is a unique active equilibrium outcome. Silence is followed by the planned pace ($a_S=0$), and a warning is followed by acceleration with probability $x$, where
\begin{equation}
x=\begin{cases}
0, & q\le u,\\[2pt]
\min\{1,b/b^\dagger\}, & q>u,
\end{cases}
\quad
b^\dagger=K_H+\dfrac{(1-q)u(K_H-K_L)}{q-u}\quad(q>u).
\label{eq:x}
\end{equation}
In the active equilibrium, both progress types report with strictly positive probability, $z_t=(b-K_tx)/C$, and $B$'s posterior belief after a warning is given by \eqref{eq:posterior}, with $m(x)=u$ whenever $0<x<1$. An additional no-report equilibrium outcome exists if and only if $b\le K_L$. All uniqueness statements hold up to indifferent cost types of measure zero.
\end{theorem}

Appendix~\ref{app:main} proves Theorem~\ref{th:main} without restricting $a_S$ ex ante; $a_S=0$ emerges endogenously from $\bar D<0$ and the law of total expectation. To interpret the threshold, suppose positive evaluations were sampled at random for mandatory publication. Observing such a report, $B$ would form posterior $q$ and earn net pursuit payoff $(v_H-v_L)(q-u)$. The sign of $q-u$ likewise governs whether \emph{voluntary} warnings trigger pursuit, whereas strategic withholding governs how often pursuit occurs. When $q>u$ and $b<b^\dagger$, full pursuit ($x=1$) would drive so many advanced laboratories into silence that $m(1)$ would fall below $u$, making acceleration unprofitable. Equilibrium instead requires an interior pursuit probability $x=b/b^\dagger$ that leaves $B$ indifferent ($m(x)=u$). Only when $b\ge b^\dagger$ does disclosure pay well enough to retain sufficiently many advanced reporters under certain pursuit.

This characterization separates the two informational channels of a warning. Holding progress beliefs fixed, certifying $r=1$ strictly raises the marginal risk cost of accelerating by $h$ relative to any prior with $p<1$. Pursuit ($x>0$) therefore arises only when favorable news about contestable progress ($q>u$) overcomes the confirmed hazard; in the benchmark limit $q\to q_0$, $D_1=D_0-h<0$ implies $q_0<u$ and $x=0$.

When $b>K_L$, low-progress senders with $k$ near zero strictly prefer to warn even if $a_W=1$, ruling out complete silence. When $b\le K_L$, PBE admits a silent outcome supported by pessimistic off-path beliefs ($\Prob(H\mid W)=1$, inducing $a_W=1$). Standard signaling refinements from \citet{chokreps} distinguish the stability of this silent trap (proof in Appendix~\ref{app:main}).

\begin{corollary}[Off-path belief refinements]\label{cor:refinement}
Suppose $b\le K_L$. The no-report equilibrium survives the Intuitive Criterion, but fails Criterion D1; hence Criterion D1 uniquely selects the active equilibrium outcome across the entire parameter domain.
\end{corollary}

Because low-cost laboratories of \emph{both} progress types ($k<b$) strictly profit from deviating to $W$ if $a_W=0$, neither progress type is equilibrium-dominated, leaving the Intuitive Criterion without bite. Under Criterion D1, however, any advanced sender $(H,k)$ with $k\le b$ weakly gains only when $a_W\le (b-k)/K_H\le b/K_H$, whereas a low-progress sender $(L,k')$ with small cost $k'>0$ strictly gains on the strictly larger interval $a_W\in[0,(b-k')/K_L)$, forcing $\Prob(H\mid W)=0$ and unraveling silence.

\subsection{Selection, reporting subsidies, and what each prediction requires}
\begin{proposition}[Selection and reporting rewards]\label{prop:selection}
In the active equilibrium, $x>0$ implies $z_H<z_L$ and $m(x)<q$. Holding $(p,q,q_0,v_H,v_L,h,c,K_H,K_L,C)$ fixed and raising $b$ within $(0,C-K_H)$: for $q>u$ and $b<b^\dagger$, both reporting rates $z_H,z_L$ and the conditional pursuit probability $x$ strictly increase, while the unconditional pursuit probability $\Prob(a=1)$ grows proportionally to $b^2$, even though the warning posterior $m(x)=u$ remains constant; for $b\ge b^\dagger$, $x=1$ and $\Prob(a=1)$ grows linearly with $b$.
\end{proposition}

\begin{proof}
Whenever $x>0$, $z_L-z_H=(K_H-K_L)x/C>0$, so Bayes' rule gives $m(x)<q$. For $q>u$ and $b<b^\dagger$, substituting $x=b/b^\dagger$ yields
\[
 z_t=\frac{b}{C}\bigl(1-K_t/b^\dagger\bigr),\qquad
 \Prob(a=1)=p\,[qz_H+(1-q)z_L]\,x,
\]
while $m(x)=u$ by Theorem~\ref{th:main}. Since $b^\dagger>K_H>K_L$, each $z_t$ and $x$ increases linearly in $b$, making $\Prob(a=1)$ proportional to $b^2$. For $b\ge b^\dagger$, $x=1$ and $z_t=(b-K_t)/C$, agreeing at $b=b^\dagger$.
\end{proof}

A reporting subsidy thus operates on two margins that offset each other in beliefs while compounding in pursuit. Holding $x$ fixed, raising $b$ brings previously silent projects---disproportionately advanced ones in proportional terms---into the reporting pool, which would lift $m(x)$ above $u$. In the mixed equilibrium, $B$ instead accelerates more frequently after each warning, deterring just enough advanced entrants to keep $m(x)$ pinned at $u$. Consequently, an analyst who evaluates a subsidy by checking whether published warnings remain technically valid or whether the average progress of reporters has shifted would find no change, yet competitive pursuit per warning and overall race frequency have strictly intensified.

It is important to distinguish the assumptions underlying these two predictions, both of which are analyzed formally in Appendix~\ref{app:cost}:
\begin{itemize}
\item \emph{Co-movement of reporting and pursuit.} The conclusion that a higher subsidy strictly increases $z_H$, $z_L$, and $x$ relies on asymmetric pursuit losses ($K_H>K_L$) and log-concavity of the sender's cost distribution $F$, without requiring uniform costs or homogeneous rivals (Appendix~\ref{app:sender_cost}--\ref{app:rival_cost}).
\item \emph{Exact invariance versus attenuation of the warning posterior.} Exact constancy of $m(x)=u$ in the mixed region comes from the competitor facing a single pursuit threshold $u$, which requires exact indifference whenever $0<x<1$. If different competitors instead face heterogeneous private acceleration costs---reflecting, for instance, varying compute slack or opportunity costs---pursuit probability rises smoothly with the posterior belief (Appendix~\ref{app:rival_cost}). Raising $b$ then strictly increases $z_H$, $z_L$, $x$, \emph{and} the warning posterior, but endogenous rival acceleration still strictly attenuates the rise in $m$ relative to a fixed-pursuit benchmark, converging back to exact posterior invariance as rival cost dispersion shrinks.
\end{itemize}

\section{Accident risk, optimal subsidies, and reporting channels}\label{sec:safety}
To evaluate safety implications, interpret the hazard as a single industry-wide accident to which both firms are exposed. Let $\pi_r^0$ denote the baseline accident probability in hazard state $r$, and suppose post-decision accident risk is
\begin{equation}
 \pi(r,a,W)=\pi_r^0+d_r a-\sigma\ind_{\{W\}},
 \qquad d_1>d_0\ge 0,\quad \sigma>0, \label{eq:pi}
\end{equation}
where parameters keep $\pi(r,a,W)\in[0,1]$. Defensive action triggered by $W$ lowers accident probability by $\sigma$, whereas accelerating research raises it by $d_r$. If $B$ suffers loss $L_B>0$ in an accident, \eqref{eq:D} arises with $h=L_B(d_1-d_0)$ and $c$ equal to $B$'s direct acceleration cost plus $L_Bd_0$. Similarly, let $L_A>0$ denote $A$'s accident loss, $s\ge0$ an external private reward or subsidy for reporting, and $J_t\ge0$ ($J_H>J_L$) $A$'s pure competitive loss from pursuit net of accident risk. Then \eqref{eq:A} holds with
\begin{equation}
 b=L_A\sigma+s,\qquad K_t=J_t+L_A d_1, \label{eq:micro}
\end{equation}
which remains within the maintained support bound $C>b+K_H$ whenever $0\le s<\bar s\equiv C-K_H-L_A\sigma$. A subsidy changes $s$ while leaving mitigation $\sigma$ fixed, whereas better defensive infrastructure raises $\sigma$ and $b$ together.

\begin{proposition}[What voluntary reporting reveals about safety]\label{prop:safety}
Let $\bar z=q z_H+(1-q)z_L$ in the active equilibrium. Relative to a benchmark in which warnings cannot be sent, the change in the ex ante accident probability is
\begin{equation}
 \Delta \Prob(\mathrm{accident})=p\bar z\,(d_1x-\sigma). \label{eq:change}
\end{equation}
Every reporting sender's participation constraint implies
\begin{equation}
 L_A(\sigma-d_1x)\ \ge\ k+J_t x-s. \label{eq:bound}
\end{equation}
Consequently, when $s=0$, active voluntary disclosure strictly reduces the ex ante accident probability. When $0<\sigma<d_1$ and $q>u$, active disclosure increases accident risk relative to no reporting if and only if $b>b^\dagger\sigma/d_1$.
\end{proposition}

\begin{proof}
Without a reporting channel, $\bar D<0$ implies $a=0$ always. Under active disclosure, risk changes only when $r=1$ and $A$ sends $W$ (probability $p\bar z$), subtracting $\sigma$ and adding $d_1$ with probability $x$, which gives \eqref{eq:change}. Substituting \eqref{eq:micro} into $b-k\ge K_tx$ yields \eqref{eq:bound}. At $s=0$, the right-hand side of \eqref{eq:bound} is strictly positive for almost all reporters ($k>0$ almost surely), forcing $\sigma-d_1x>0$ whenever $\bar z>0$. Substituting $x=\min\{1,b/b^\dagger\}$ into \eqref{eq:change} with $\sigma<d_1$ gives $d_1x-\sigma>0$ if and only if $b>b^\dagger\sigma/d_1$.
\end{proof}

Proposition~\ref{prop:safety} is a revealed-preference restriction: absent external rewards ($s=0$), $A$'s willingness to absorb verification costs $k$ and competitive losses $J_tx$ certifies that mitigation outweighs the expected accident damage from induced pursuit. Publicity, compliance safe harbors, or direct subsidies ($s>0$) relax this bound, as would asymmetric externalities in which a warning protects $A$'s own users while pursuit endangers external parties uninternalized by $L_A$.

Within the maintained parameter domain, Propositions~\ref{prop:selection} and~\ref{prop:safety} pin down the accident-minimizing reporting subsidy.

\begin{corollary}[Risk-minimizing subsidy on the admissible domain]\label{cor:subsidy}
Suppose $q>u$, $0<\sigma<d_1$, and $C>\max\{L_A\sigma,b^*\}+K_H$, where $b^*\equiv \sigma b^\dagger/(2d_1)$ and $\bar s\equiv C-K_H-L_A\sigma>0$. Across all admissible subsidies $s\in[0,\bar s)$ along the active equilibrium branch, the ex ante accident probability is uniquely minimized at
\begin{equation}
 s^*=\max\left\{0,\;\sigma\Bigl(\frac{b^\dagger}{2d_1}-L_A\Bigr)\right\}\in[0,\bar s). \label{eq:opt_s}
\end{equation}
If $b^\dagger>2L_Ad_1$, a strictly positive subsidy $s^*>0$ minimizes accident risk (yielding $x(b^*)=\sigma/(2d_1)<1/2$); if $b^\dagger\le 2L_Ad_1$, the unique minimizer on $[0,\bar s)$ is $s^*=0$, and any positive admissible subsidy strictly raises accident risk.
\end{corollary}

\begin{proof}
Let $\bar K=qK_H+(1-q)K_L<K_H<b^\dagger$. By hypothesis, $b=L_A\sigma+s$ ranges over $[L_A\sigma,C-K_H)$, which contains $b^*$. For $b\in(0,\min\{b^\dagger,C-K_H\})$, substituting $x=b/b^\dagger$ and $\bar z=(b/C)(1-\bar K/b^\dagger)$ into \eqref{eq:change} yields the strictly convex quadratic
\[
 \Delta \Prob(\mathrm{accident})=\frac{p}{C}\Bigl(1-\frac{\bar K}{b^\dagger}\Bigr)\,b\Bigl(\frac{d_1}{b^\dagger}b-\sigma\Bigr),
\]
whereas for any admissible $b\ge b^\dagger$, $x=1$ implies $\Delta \Prob(\mathrm{accident})=p\bar z(d_1-\sigma)>0$. Since $0<\sigma<d_1$, $b^*=\sigma b^\dagger/(2d_1)<b^\dagger/2$ uniquely minimizes $\Delta\Prob(\mathrm{accident})$ over $b\in(0,C-K_H)$. Substituting $b=L_A\sigma+s$ with $s\in[0,\bar s)$ gives \eqref{eq:opt_s}.
\end{proof}

When $A$'s private accident stake $L_A$ is small relative to $b^\dagger/(2d_1)$, laissez-faire under-provides warnings because $A$ does not internalize $B$'s losses or broader harm, justifying a bounded positive subsidy $s^*>0$. Beyond $b^*$, however, rising pursuit erodes safety faster than extra disclosures improve it, turning public warnings net risk-increasing once $b>2b^*=b^\dagger\sigma/d_1$ (when admissible).\footnote{Equation~\eqref{eq:change} measures physical accident probability; ranking social surplus would additionally weight the value of accelerated innovation against verification costs $k$ and the marginal cost of public funds.}

\subsection{Confidential versus public reporting channels}
Suppose $A$ can instead transmit verified findings to defenders through a confidential channel unobserved by $B$. Consider first a pure institutional comparison in which confidential reporting is the \emph{only} channel and never leaks. Then $B$ retains its prior ($a=0$). With identical cost $k$, reward, and defensive reach $\sigma$, positive types report whenever $k\le b$ (rate $b/C$), so active public disclosure generates excess accident probability
\[
 p\sigma(b/C-\bar z)+p\bar z\, d_1x ,
\]
reflecting both lost precautions from withholding ($b/C-\bar z$) and pursuit-induced risk ($\bar z d_1x$). If a standalone confidential channel instead reaches fewer defenders, delivering smaller mitigation $\sigma_c<\sigma$ and reporting rate $z_c$, public-only disclosure is safer precisely when $\bar z(\sigma-d_1x)>z_c\sigma_c$.

In practice, frontier laboratories often have simultaneous access to coordinated confidential reporting and public announcements. Suppose that when $r=1$, $A$ chooses among silence $S$, a confidential report $W_c$ unobserved by $B$ that yields gross private return $b_c\in(0,b)$ at the same verification cost $k$, and a public warning $W_p$ observed by $B$ that yields gross return $b$ (with $C>b+K_H$). Write $\Delta b\equiv b-b_c>0$, let $x_{\mathrm{pub}}=\min\{1,b/b^\dagger\}$ denote the public-only pursuit rate when $q>u$, and call an equilibrium \emph{publicly active} if $\Prob(W_p)>0$. Because $B$'s pursuit decision adjusts endogenously to channel sorting, introducing $W_c$ alters pursuit frequency rather than pushing the public-warning posterior below $u$ (proof in Appendix~\ref{app:main}).

\begin{proposition}[Coexisting public and confidential channels]\label{prop:channels}
Maintain \eqref{eq:domain} and $0<b_c<b<C-K_H$. In every publicly active equilibrium, absence of a public warning is followed by the planned pace. For $q\le u$, or for $q>u$ and $\Delta b>K_H x_{\mathrm{pub}}$, the confidential channel is unused and the unique publicly active outcome coincides with Theorem~\ref{th:main}. For $q>u$ and $0<\Delta b<K_H x_{\mathrm{pub}}$, there is a unique publicly active equilibrium outcome:
\begin{enumerate}[label=(\roman*),nosep,topsep=3pt]
\item A public warning triggers pursuit with probability $x_c=\Delta b/K_H < x_{\mathrm{pub}}$, and $B$'s posterior belief after $W_p$ is $\Prob(H\mid W_p)=u$.
\item Low-progress senders report exclusively in public ($z_L^p=(b-K_Lx_c)/C$, $z_L^c=0$), while advanced senders with $k\le b_c$ split between public warnings ($z_H^p=\lambda b_c/C$) and confidential reports ($z_H^c=(1-\lambda)b_c/C$), where $\lambda=\frac{u(1-q)}{q(1-u)}\frac{b-K_Lx_c}{b_c}\in(0,1)$.
\end{enumerate}
\end{proposition}

Comparing Proposition~\ref{prop:channels} with the mixed public-only equilibrium ($q>u$, $b\le b^\dagger$) underscores the equilibrium role of rival pursuit. Advanced laboratories indeed sort into the confidential channel, whereas low-progress reporters stay public. Yet that migration does \emph{not} depress the progress posterior $\Prob(H\mid W_p)$ below $u$: if it did, $B$ would stop pursuing ($x=0$), eliminating any reason to sacrifice $\Delta b$ for confidentiality. Instead, $\Prob(H\mid W_p)$ remains fixed at $u$ while per-warning pursuit drops from $b/b^\dagger$ to $\Delta b/K_H$, raising total hazard reporting for both types. Moreover, if both channels receive the same external reward $s$---so $b=L_A\sigma+s$ and $b_c=L_A\sigma_c+s$ with $0<\sigma_c<\sigma$---then $\Delta b=L_A(\sigma-\sigma_c)$ is independent of $s$, capping pursuit at $x_c=L_A(\sigma-\sigma_c)/(J_H+L_Ad_1)\le (\sigma-\sigma_c)/d_1$ and guaranteeing $\sigma-d_1x_c\ge\sigma_c>0$ even under large subsidies.

\subsection{A numerical illustration}
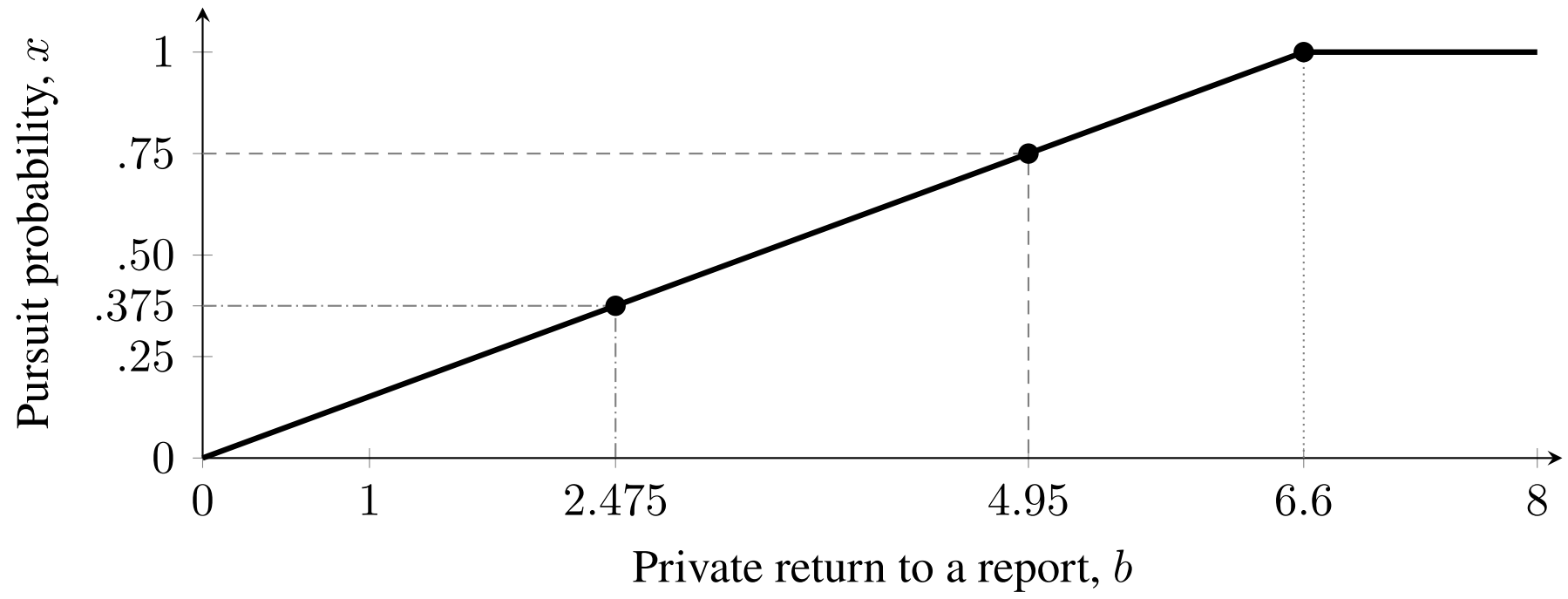
\begin{figure}[htbp]\centering
\begin{tikzpicture}
\begin{axis}[width=119mm,height=5.0cm,axis lines=left,xmin=0,xmax=8.15,ymin=0,ymax=1.11,xlabel={Private return to a report, $b$},ylabel={Pursuit probability, $x$},xtick={0,1,2.475,4.95,6.6,8},xticklabels={$0$,$1$,$2.475$,$4.95$,$6.6$,$8$},ytick={0,.25,.375,.5,.75,1},yticklabels={$0$,$.25$,$.375$,$.50$,$.75$,$1$},tick label style={font=\small},label style={font=\small},clip=false]
\addplot[gray,densely dashdotted,thin] coordinates{(0,.375)(2.475,.375)(2.475,0)};
\addplot[gray,densely dashed,thin] coordinates{(0,.75)(4.95,.75)(4.95,0)};
\addplot[gray,densely dotted,thin] coordinates{(6.6,0)(6.6,1)};
\addplot[black,very thick] coordinates{(0,0)(6.6,1)(8,1)};
\addplot[only marks,mark=*,mark size=2pt,black] coordinates{(2.475,.375)(4.95,.75)(6.6,1)};
\end{axis}
\end{tikzpicture}
\caption{Pursuit after a warning in the public-only active equilibrium. Accident risk is minimized at $b^*=2.475$ ($x=.375$, dash-dotted guides), breaks even with no reporting at $b=4.95$ ($x=.75$, dashed guides), and reaches certain pursuit above $b^\dagger=6.6$ (dotted line). For $0<b\le1$, an unrefined silent equilibrium coexists with the active branch. Parameters follow Table~\ref{tab:example}.}\label{fig:example}
\end{figure}

Set $p=.2$, $q=.7$, $q_0=.1$, $v_H=10$, $v_L=0$, $c=h=3$, $K_H=3$, $K_L=1$, and $C=12$, with risk parameters $d_1=.04$ and $\sigma=.03$. Here $B$'s prior return to accelerating is $-1.4$, whereas confirming $r=1$ without selection raises the return to $+1$, giving $u=.6$ and $b^\dagger=6.6$. Public-only warnings reduce accident risk whenever $x<.75$ ($b<4.95$), and Corollary~\ref{cor:subsidy} places the risk-minimizing return at $b^*=2.475$ ($x^*=.375$, admissible since $C-K_H=9$). Table~\ref{tab:example} contrasts $b=4$ and $b=7$ under public-only reporting, and Fig.~\ref{fig:example} plots equilibrium pursuit across $b$.

\begin{table}[htbp]
\centering\caption{Reporting and pursuit under public-only disclosure as the private return rises}\label{tab:example}
\begin{tabular}{@{}lrr@{}}
\toprule
 & $b=4$ & $b=7$\\ \midrule
Advanced laboratories reporting, among positive evaluations & 18.18\%&33.33\%\\
Other laboratories reporting, among positive evaluations &28.28\%&50.00\%\\
Probability of advanced progress given a warning &60.00\%&60.87\%\\
Probability of pursuit given a warning &60.61\%&100.00\%\\
Probability of a public warning, all projects &4.24\%&7.67\%\\
Change in accident probability, percentage points &$-0.0244$&$+0.0767$\\\bottomrule
\end{tabular}
\parbox{.94\linewidth}{\small\textit{Notes.} The two columns vary only the external reporting reward $s$. Underlying primitives are $L_A=25$, $L_B=100$, $d_0=.01$, $J_L=0$, $J_H=2$, direct acceleration cost $2$, $\pi_0^0=.02$, and $\pi_1^0=.10$, implying $L_A\sigma=.75$, admissible bound $\bar s=8.25$, and rewards $s=3.25$ (at $b=4$) and $s=6.25$ (at $b=7$). Entries are exact model calculations.}
\end{table}

At $b=4$, a sender with cost $k=3$ reports if $t=L$ (cutoff $3.3939$) but stays silent if $t=H$ (cutoff $2.1818$) despite holding identical hazard evidence. Varying $s$ traces the quadratic safety profile: at $s=0$ ($b=.75$), active disclosure induces pursuit at rate $.75/6.6\approx .114$ and lowers accident risk by $.0202$ percentage points (coexisting with an unrefined silent PBE since $b<K_L$); the optimal subsidy $s^*=1.725$ ($b^*=2.475$) deepens the risk reduction to $-.0394$ percentage points; raising rewards to $b=4$ erodes the gain back to $-.0244$ percentage points, and exceeding $b=4.95$ turns public warnings net risk-increasing.

Now introduce a narrower confidential channel alongside public warnings at the $b=4$ ($s=3.25$) primitives, setting $\sigma_c=.02$ so $b_c=25(.02)+3.25=3.75$ and $\Delta b=.25<K_Hx_{\mathrm{pub}}\approx 1.818$. By Proposition~\ref{prop:channels}, adding the confidential option leaves the posterior after a public warning unchanged at $60.00\%$, while slashing pursuit per public warning from $60.61\%$ to $x_c=.25/3\approx 8.33\%$. Low-progress positive types report publicly at rate $32.64\%$, and advanced positive types report at combined rate $31.25\%$ (with $\lambda=47/70\approx 67.14\%$ choosing $W_p$ and $32.86\%$ choosing $W_c$), expanding the ex ante accident reduction more than sixfold from $-.0244$ to $-.1593$ percentage points.

\section{Implications for evidence and interpretation}
Our results carry three lessons for empirical and policy evaluation of safety disclosures. First, observing stable technical quality or progress content across published warnings after a policy shift does not imply that competitive incentives are unchanged: in our mixed equilibria, subsidies and confidential channels operate primarily through the rival's conditional pursuit rate and the composition of unpublished findings. Auditing selection directly therefore requires pairing voluntary public warnings with confidential mandatory evaluation logs held by an oversight body or third-party evaluator.

Second, institutional shifts between compulsory and voluntary reporting identify selection when orthogonal to underlying technical progress. For $q>u$ and $b<b^\dagger$, a random positive report under mandatory auditing implies posterior $q>u$ and strict pursuit ($x=1$), whereas the same verifiable finding disclosed voluntarily yields posterior $u$ (or an attenuated posterior under heterogeneous rival costs) and probabilistic pursuit ($x<1$). If mandatory rules also alter which internal evaluations firms run, observed differences additionally reflect endogenous information acquisition \citep{friehe}.

Third, empirically coding a competitor's post-warning response as risk-increasing pursuit ($a=1$) requires comparing action dates against preexisting compute or deployment schedules, and separating accelerated capability pushes from purely defensive access expansions that do not carry the hazard cost $h$ in \eqref{eq:D}.

\section{Conclusion}
A verifiable safety warning about dual-use technology reaches both defenders who mitigate harm and competitors who decide whether to accelerate their own research. Because rival pursuit is costlier to laboratories with valuable contested leads, advanced developers withhold true hazard findings more frequently, leading rational competitors to discount the progress implied by voluntary warnings.

Accounting for both sides' equilibrium adjustments is essential for policy design. Raising reporting rewards increases report volume and per-warning pursuit simultaneously while leaving the inferred progress of public reporters unchanged (or attenuated under rival cost heterogeneity), whereas offering an equally rewarded confidential channel alongside public warnings curbs pursuit by giving advanced reporters an exit option. Absent external reporting rewards, revealed preference ensures that voluntary warnings lower expected accident risk despite provoking pursuit; when external subsidies are used to draw out withheld reports, safety is maximized by keeping those rewards strictly bounded.

\appendix
\section{Equilibrium proofs and channel coexistence}\label{app:main}
\begin{proof}[Proof of Theorem~\ref{th:main} and Corollary~\ref{cor:refinement}]
Write $a_W,a_S\in[0,1]$ for $B$'s behavioral strategies. For a sender with $r=1$, reporting rather than remaining silent yields net payoff $b-k-K_t(a_W-a_S)$, so $z_t(y)=\Pos{b-K_ty}/C$ with $y=a_W-a_S$. Because $C>b+K_H$, $z_t<1$ for all $y\in[-1,1]$, ensuring that $S$ always has strictly positive probability.

\emph{Step 1. Behavior after silence in active equilibria.}
Suppose $y<0$. Then $z_H>z_L>0$, giving $m(y)>q$. If $q>u$, $B$ strictly prefers acceleration after $W$ ($a_W=1$), contradicting $y<0$. If $q\le u$, advanced projects report at a higher rate than low-progress projects among $r=1$ states, so the share of $t=H$ among silent $r=1$ states is strictly below $q\le u$. Pursuit therefore yields a strictly negative expected return conditional on $(r=1,S)$ and $D_0<0$ conditional on $(r=0,S)$, implying $a_S=0$ and again contradicting $y<0$.

Suppose $y=0$. In an active equilibrium, $z_H=z_L=b/C>0$ and $m(0)=q$. If $q\le u$, pursuit after $S$ yields $D_0<0$ on $(r=0,S)$ and $D_1\le 0$ on $(r=1,S)$, forcing $a_S=0$ and hence $a_W=0$. If $q>u$, $m(0)=q>u$ requires $a_W=1$ and thus $a_S=1$, contradicting $\bar D<0$ by the law of total expectation.

Finally, suppose $y>0$. Then $a_W>0$, so $B$'s expected return from accelerating after $W$ must be nonnegative. Since the prior expected return $\bar D<0$ is a convex combination of the expected returns after $W$ and after $S$, the expected return after $S$ must be strictly negative, giving $a_S=0$. Hence $a_S=0$ and $y=x=a_W$ in every active equilibrium.

\emph{Step 2. Active equilibrium.}
For $x>0$, $z_H\le z_L$. If $z_H=0<z_L$, then $m(x)=0<u$, which implies $a_W=0$, a contradiction. Thus both progress types report with positive probability in any active equilibrium, with posterior odds
\begin{equation}
 \frac{m(x)}{1-m(x)}
   =\frac{q}{1-q}\frac{b-K_Hx}{b-K_Lx}. \label{eq:odds}
\end{equation}
The right-hand side of \eqref{eq:odds} strictly decreases in $x$ whenever $b-K_Hx>0$. If $q\le u$, any $x>0$ implies $m(x)<q\le u$, leaving $x=0$ as the unique solution. If $q>u$, setting $m(x_0)=u$ yields the unique crossing $x_0=b/b^\dagger>0$. Because $b^\dagger>K_H$, $x_0<b/K_H$, so $z_H(x_0),z_L(x_0)>0$. When $x_0<1$, $m(x)>u$ for $x<x_0$ (requiring $a_W=1$) and $m(x)<u$ for $x>x_0$ (requiring $a_W=0$), establishing $x=x_0$ as the unique equilibrium response. When $x_0\ge 1$, $m(x)>u$ for all $x<1$ and $m(1)\ge u$, uniquely yielding $x=1$. Optimality of $a_S=0$ follows from $\bar D<0$ for $x>0$ (and directly from $D_0<0,D_1\le0$ for $x=0$). Conditional on $S$, the posterior weights on $(1,H),(1,L),(0,H),(0,L)$ are $pq(1-z_H)$, $p(1-q)(1-z_L)$, $(1-p)q_0$, and $(1-p)(1-q_0)$ normalized by $1-p[qz_H+(1-q)z_L]$, with truncated uniform cost posteriors on $[b-K_tx,C]$ in positive states.

\emph{Step 3. Complete silence and off-path refinements.}
In a no-report equilibrium, $S$ occurs with probability one, leaving $B$ at its prior ($\bar D<0$) and forcing $a_S=0$. Deterring almost all $(1,L,k)$ senders requires $b-k-K_La_W\le 0$ for all $k>0$, hence $b\le K_La_W\le K_L$. Conversely, when $b\le K_L$, assigning off-path belief $\Prob(H\mid W)=1$ (with $r=1$ by verifiability) makes $a_W=1$ sequentially rational since $u<1$, deterring every sender with $k>0$ (any off-path belief supporting $a_W\ge b/K_L$ suffices).

To verify Corollary~\ref{cor:refinement} for $b\le K_L$, note that $B$'s set of mixed best responses across all off-path beliefs $\mu=\Prob(H\mid W)\in[0,1]$ (given $r=1$) is $a_W\in[0,1]$. Equilibrium payoff in the silent outcome is $0$ for all $r=1$ types. At $a_W=0$, deviating to $W$ yields $b-k>0$ for every $(1,t,k)$ with $k\in[0,b)$ and $t\in\{L,H\}$; thus neither progress type is equilibrium-dominated, and $\mu=1$ survives the Intuitive Criterion. Under Criterion D1, fix any advanced type $\theta_H=(1,H,k)$. If $k>b$, $\theta_H$ strictly loses from $W$ for all $a_W\in[0,1]$. If $k\in[0,b]$, $\theta_H$ weakly gains on $[0,(b-k)/K_H]\subseteq[0,b/K_H]$. Choosing $k'\in\bigl(0,b(1-K_L/K_H)\bigr)$, since $k'>0$ and $b\le K_L$, we have $b/K_H<(b-k')/K_L<1$, so $\theta_L=(1,L,k')$ \emph{strictly} gains on $[0,(b-k')/K_L)\supsetneq[0,b/K_H]$. Criterion D1 therefore eliminates $\theta_H$ for all $k\in[0,C]$, forcing $\Prob(H\mid W)=0<u$ and $a_W=0$, unraveling silence.
\end{proof}

\begin{proof}[Proof of Proposition~\ref{prop:channels}]
Let $a_p,a_0\in[0,1]$ denote $B$'s pursuit probabilities after observing $W_p$ and after observing no public warning (pooling $r=0$, $(r=1,S)$, and $(r=1,W_c)$), and set $y=a_p-a_0$. In state $r=1$, $A$'s payoffs from $S$, $W_c$, and $W_p$ are $-K_ta_0$, $b_c-k-K_ta_0$, and $(b_c-k-K_ta_0)+(\Delta b-K_ty)$, respectively. Because $C>b+K_H$, non-public states always have strictly positive probability. If $y\le 0$, then $\Delta b-K_ty\ge \Delta b>0$, so $W_p$ strictly dominates $W_c$ for all senders; $W_c$ is unused and Step~1 of Theorem~\ref{th:main} rules out $y<0$ and forces $a_0=a_p=0$ when $y=0$ (requiring $q\le u$). If $y>0$, then $a_p>0$ implies $\mathbb{E}[D_{rt}\mid W_p]\ge 0$, so $\bar D<0$ and the law of total expectation force $\mathbb{E}[D_{rt}\mid \neg W_p]<0$ and hence $a_0=0$. Thus $a_0=0$ and $y=a_p=x$ in every publicly active equilibrium.

With $a_0=0$, for any $x\ge 0$: if $q\le u$, $\Prob(H\mid W_p)\le m(x)\le q\le u$ whenever $\Prob(W_p)>0$ (since $K_H>K_L$ ensures advanced senders never favor $W_p$ over $W_c$ more strongly than low-progress senders), forcing $x=0$ and zero confidential reporting ($W_p$ strictly dominates $W_c$ at $x=0$). Now suppose $q>u$. If $\Delta b>K_Hx_{\mathrm{pub}}$, then for all $x\le x_{\mathrm{pub}}$ we have $\Delta b-K_Hx>0$, so $W_c$ is unused; and any $x>x_{\mathrm{pub}}$ (feasible only if $x_{\mathrm{pub}}<1$, i.e., $b<b^\dagger$) yields $\Prob(H\mid W_p)\le m(x)<m(x_{\mathrm{pub}})=u$, contradicting $x>0$. Hence $x=x_{\mathrm{pub}}$ uniquely. Finally, suppose $0<\Delta b<K_Hx_{\mathrm{pub}}$, so $x_c\equiv\Delta b/K_H\in(0,x_{\mathrm{pub}})\subseteq(0,1)$. Any $x<x_c$ makes $\Delta b-K_Hx>0$, leaving $W_c$ unused and giving $\Prob(H\mid W_p)=m(x)>m(x_c)>u$ (as $x_c<b/b^\dagger$), which requires $x=1>x_c$, a contradiction. Any $x>x_c$ makes $\Delta b-K_Hx<0$, so $z_H^p=0$ and $\Prob(H\mid W_p)=0<u$, requiring $x=0<x_c$, a contradiction. Thus $x=x_c$ uniquely. At $x_c$, $\Delta b-K_Lx_c=\Delta b(1-K_L/K_H)>0$, so low types report publicly iff $k\le b-K_Lx_c$ ($z_L^p=(b-K_Lx_c)/C$, $z_L^c=0$), while high types with $k\le b_c=b-K_Hx_c$ are indifferent between $W_p$ and $W_c$. Since $x_c<b/b^\dagger$, $\frac{q}{1-q}\frac{b_c}{b-K_Lx_c}>\frac{u}{1-u}$; equating posterior odds $\frac{q}{1-q}\frac{\lambda b_c}{b-K_Lx_c}$ to $\frac{u}{1-u}$ uniquely pins down $\lambda\in(0,1)$ and verifies $B$'s indifference at $\Prob(H\mid W_p)=u$.\footnote{At the knife-edge $\Delta b=K_Hx_{\mathrm{pub}}$, $x=x_{\mathrm{pub}}$ remains the unique pursuit rate: if $b\le b^\dagger$, $\lambda=1$ ($z_H^c=0$) uniquely maintains $\Prob(H\mid W_p)=u$, whereas if $b>b^\dagger$ ($x_{\mathrm{pub}}=1$), any $\lambda\in[\frac{u(1-q)}{q(1-u)}\frac{b-K_L}{b_c},1]$ keeps $\Prob(H\mid W_p)\ge u$. An unrefined equilibrium with $\Prob(W_p)=0$ (all $k\le b_c$ using $W_c$) exists iff $\Delta b\le K_L$.}
\end{proof}

\section{Robustness: sender cost distributions and rival heterogeneity}\label{app:cost}

\subsection{Non-uniform sender reporting costs}\label{app:sender_cost}
Let $k$ follow a continuous distribution $F$ supported on $[0,C]$, differentiable with strictly positive density on $(0,C)$, satisfying $F(0)=0$ and $C>b+K_H$. Assume $\log F$ is concave on $(0,C]$, and set $z_t(y)=F(b-K_ty)$ (with $F(w)=0$ for $w\le 0$). Steps~1 and~3 of Appendix~\ref{app:main} carry over verbatim because they rely only on payoff monotonicity and cost support. Writing $g(w)=(\log F)'(w)>0$ for $w\in(0,C]$, whenever $x\ge0$ and $b-K_Hx>0$, differentiating the log posterior odds with respect to $x$ gives
\[
 -K_H g(b-K_Hx)+K_L g(b-K_Lx)<0,
\]
since $b-K_Hx\le b-K_Lx$, $K_H>K_L$, and $g$ is nonincreasing. Starting from $m(0)=q$, if $b-K_Hx$ hits zero at some $\bar x\le 1$, then $m(x)\to 0$ as $x\uparrow\bar x$ while $z_L(\bar x)>0$; otherwise $m(1)$ is well defined. Hence a unique active outcome again exhibits $x=0$ for $q\le u$, $x=1$ when $q>u$ and $m(1)\ge u$, and a unique interior $x\in(0,1)$ solving $m(x)=u$ otherwise.

Moreover, if $\log F$ is \emph{strictly} concave on $(0,C]$ ($g$ strictly decreasing), Proposition~\ref{prop:selection}'s qualitative comparative statics hold without uniformity. In the mixed region ($0<x<1$), $m(x)=u$ fixes $\log F(b-K_Hx)-\log F(b-K_Lx)=\log[u(1-q)/(q(1-u))]$. Implicit differentiation with respect to $b$ yields $g_H(1-K_Hx'(b))=g_L(1-K_Lx'(b))$, where $g_t\equiv g(b-K_tx)$ satisfies $g_H>g_L>0$, giving
\[
 x'(b)=\frac{g_H-g_L}{K_H g_H-K_L g_L}>0,\qquad
 1-K_H x'(b)=\frac{(K_H-K_L)g_L}{K_H g_H-K_L g_L}>0.
\]
Since $1-K_Lx'(b)>1-K_Hx'(b)>0$, raising $b$ strictly increases $z_H$, $z_L$, $x$, and $\Prob(a=1)$ while keeping $m(x)=u$ fixed; meanwhile, the safety bound \eqref{eq:bound} is completely distribution-free.

\subsection{Heterogeneous rival acceleration costs}\label{app:rival_cost}
Retain strictly log-concave sender costs $F$ as in Appendix~\ref{app:sender_cost}, and suppose Laboratory $B$ privately draws its cost $c$ in \eqref{eq:D} from a distribution $G$ on $[\underline c,\overline c]$, independent of $(r,t,k)$, differentiable with $G'>0$ on $(\underline c,\overline c)$. Maintain $D_0(\underline c)<0$, $\bar D(\underline c)<0$, and $0<u(\underline c)<1$, where $u(c)\equiv (c+h-v_L)/(v_H-v_L)$. Write $x=\int a_W(c)\,dG(c)$, $x_S=\int a_S(c)\,dG(c)$, and $y=x-x_S$.

In any active equilibrium, $x_S=0$: if $y>0$, then $x>0$ implies $\mathbb{E}[D_{rt}(\underline c)\mid W]\ge 0$, so $\bar D(\underline c)<0$ forces $\mathbb{E}[D_{rt}(c)\mid S]\le \mathbb{E}[D_{rt}(\underline c)\mid S]<0$ for all $c$; if $y=0$, $m(0)=q$, so either $q\le u(\underline c)$ (giving $x_S=x=0$) or $q>u(\underline c)$ (giving $x_S=x>0$, which requires $\mathbb{E}[D_{rt}(\underline c)\mid S]\ge 0$ alongside $D_1(\underline c)>0$, contradicting $\bar D(\underline c)<0$); and if $y<0$, $z_H>z_L>0$ implies $m(y)>q>\Prob(H\mid r=1,S)$, so any $c$ with $\mathbb{E}[D_{rt}(c)\mid S]\ge 0$ must satisfy $\mathbb{E}[D_{rt}(c)\mid r=1,S]>0$ (as $D_0(c)<0$) and hence $\mathbb{E}[D_{rt}(c)\mid W]>0$ ($a_W(c)=1\ge a_S(c)$), yielding $x\ge x_S$, a contradiction.

Given posterior $m$ after $W$, $B$ pursues iff $c\le v_L+(v_H-v_L)m-h$, generating aggregate pursuit probability $\psi(m)=G\bigl(v_L+(v_H-v_L)m-h\bigr)$, which is continuous and nondecreasing. Since $x\mapsto m(x,b)$ is continuous and strictly decreasing wherever $z_H(x)>0$ with $m(0,b)=q$, the fixed-point condition $x=\psi(m(x,b))$ admits a unique active equilibrium outcome: $x=0$ if $q\le u(\underline c)$, and $x>0$ if $q>u(\underline c)$. At any interior equilibrium $x(b)\in(0,1)$, $\psi'(m)=(v_H-v_L)G'>0$, $\partial m/\partial b>0$, and $\partial m/\partial x<0$. Implicitly differentiating $x=\psi(m(x,b))$ gives
\[
 x'(b)=\frac{\psi'(m)\,\frac{\partial m}{\partial b}}{1-\psi'(m)\,\frac{\partial m}{\partial x}}>0,\qquad
 \frac{d\,m(x(b),b)}{db}=\frac{\frac{\partial m}{\partial b}}{1-\psi'(m)\,\frac{\partial m}{\partial x}}\in\Bigl(0,\frac{\partial m}{\partial b}\Bigr).
\]
Furthermore, $\frac{d}{db}m(x(b),b)>0$ implies $g_H(1-K_Hx'(b))>g_L(1-K_Lx'(b))$, hence $x'(b)<(g_H-g_L)/(K_Hg_H-K_Lg_L)<1/K_H$ and $1-K_Lx'(b)>1-K_Hx'(b)>0$. Thus raising $b$ still strictly increases $z_H$, $z_L$, and $x$. With rival heterogeneity, the warning posterior rises slightly with $b$, but rival pursuit adjustment strictly attenuates that rise by the factor $[1-\psi'(m)(\partial m/\partial x)]^{-1}<1$, converging to exact posterior invariance ($dm/db\to 0$) as $G$ concentrates at a common cost ($\psi'(m)\to\infty$).

\end{document}